\documentclass[twocolumn,prd,groupedaddress,nofootinbib,showpacs]{revtex4}

\usepackage{graphicx}
\usepackage{dcolumn}
\usepackage{bm}

\begin{document}

\title{QCD/String holographic mapping and glueball mass spectrum}

\author{Henrique Boschi-Filho}
\email{boschi@if.ufrj.br}
\affiliation{Instituto de F\'{\i}sica, 
Universidade Federal do Rio de Janeiro, Caixa Postal 68528, RJ 21945-970 -- Brazil}
\author{Nelson R. F. Braga}
\email{braga@if.ufrj.br}
\affiliation{Instituto de F\'{\i}sica,
Universidade Federal do Rio de Janeiro, Caixa Postal 68528, RJ 21945-970 -- Brazil}

 
\begin{abstract} 
Recently Polchinski and Strassler reproduced the high energy QCD scaling
at fixed angles from a gauge string duality inspired in AdS/CFT correspondence.
In their approach a confining gauge theory is taken as approximately dual to 
an AdS space with an IR cut off. Considering such an approximation (AdS slice) we
found  a one to one holographic mapping between bulk and boundary scalar fields.
Associating the bulk fields with dilatons and the boundary fields with
glueballs of the confining gauge theory we  
also found the same high energy QCD scaling.
Here, using this holographic mapping we give a simple estimate for the mass ratios
of the glueballs assuming the AdS slice approximation to be valid at low energies. 
We also compare these results to those coming from 
supergravity and lattice QCD.

\end{abstract}
\pacs{ 11.25.Tq ; 12.38.Aw ; 12.39.Mk .}

\maketitle

\vfill\eject

  
Recently Polchinski and Strassler reproduced important observed properties of
QCD from string theory in AdS space\cite{PS,PS2}.  In these articles they used
a model for the dual of a confining gauge theory which is approximately
an AdS slice with an infrared cut off. Using this model they were able\cite{PS} to
obtain the high energy scaling of QCD scattering amplitudes for fixed 
angles\cite{QCD1,BRO} as well as the Regge regime. 
Further they proposed\cite{PS2} a way of analyzing the deep inelastic scattering 
and Bjorken scaling in terms of string theory.  
Using the same kind of AdS slice we proposed\cite{BB3} a one to one holographic 
mapping between low energy string dilaton states in AdS bulk and massive composite
operators on its boundary.  
From this mapping we also obtained a scaling for high energy amplitudes 
at fixed angles similar to that of QCD and of Polchinski and Strassler
(see also \cite{GI,BT,AN}). 

The gauge/string duality considered in refs. \cite{PS,PS2} 
was inspired in the AdS/CFT correspondence proposed recently 
by Maldacena\cite{Malda} where SU(N) conformal gauge theory with ${\cal N} = 4\,$ 
supersymmetry is dual to string theory in AdS space (times a compact manifold).
The prescriptions for realizing of this correspondence obtaining
boundary correlation functions in terms of bulk fields were proposed in \cite{GKP,Wi} 
(see also \cite{Malda2} for a review). 
The AdS/CFT correspondence can be understood as a realization of the holographic 
principle\cite{HOL1,HOL2,HOL3,HOL4}. 
This principle asserts that the degrees of freedom of a theory with gravity 
defined in a given space can be mapped on the corresponding boundary.

In the AdS/CFT correspondence the higher the energy of a given boundary process,
the closer to the horizon is the bulk dual. Restricting boundary process 
to energies higher than some IR cut off would then correspond to
restricting the bulk to some region in the neighborhood of the horizon.
That is a motivation for taking an AdS slice as an approximation for the space dual 
to a boundary confining gauge theory.
Such a gauge theory with an infrared cut off can be related to 
${\cal N} = 1^\ast$ supersymmetric Yang Mills theory\cite{PS,PS2}
(see also \cite{BCE}). 
This model leads to QCD like behavior at high energies.
An AdS slice was used before in \cite{RS1,RS2} to propose a solution to the
hierarchy problem. An approach based on string theory to ${\cal N} = 1$
super Yang Mills has been proposed in \cite{MN}.

An approach to QCD from AdS/CFT correspondence was 
proposed by Witten\cite{Wi2}. It consists of breaking supersymmetry
with different compactifications of AdS space. This involves at least 
one circle $S_1$ where anti periodic boundary 
conditions are assumed for the fermionic fields (the bosonic ones are periodic).
These compactifications leads to AdS-Schwarzschild black hole metrics
that can be related to QCD$_3$ or QCD$_4$\cite{Wi2}.
This approach can be used to estimate glueball masses from supergravity models relating
glueballs with the bulk dilaton modes in 
different dimensions\cite{OZ1}. This idea was implemented in \cite{MASSG}
for QCD$_3$ and QCD$_4$ where the supergravity equations with the black hole metric 
do not allow analytic solutions but the eigenvalues related to the glueball masses
can be found using a WKB method (see also 
\cite{MASSG2,MASSG3,MASSG4,MASSG5,MASSG6,MASSG7}). 

Here we use the mapping proposed in \cite{BB3} between bulk dilatons and 
massive boundary operators defined in the AdS slice
to estimate in a simple way  the ratio of boundary masses. 
This slice has an infrared cut off that Polchinski and Strassler identified
with the mass of the lightest glueball. Using this identification
and assuming that the approximated duality is still valid for low energies
we identify our boundary operators with glueballs of the confining gauge theory.

The general structure of the holographic mapping was taken from ref.(\cite{BB2})
where we introduced a mapping between scalar fields in AdS bulk and boundary.
Using this mapping and low energy string theory approximation 
we found the QCD like scaling for high energy amplitudes\cite{BB3}.

We  consider an AdS$_5 \times$ S$^5$ space
with radius $R$ described by Poincar\'e coordinates
\begin{equation}
\label{metric}
ds^2=\frac {R^2 }{ z^2}\Big( dz^2 \,+(d\vec x)^2\,
- dt^2 \Big) \,+ \,R^2 d\Omega_5^2 \,\,,
 \end{equation}

\noindent where $\Omega_5$ corresponds to the five dimensional sphere
$S^5$.  According to the AdS/CFT correspondence, glueballs are related to
closed strings. At energies much lower than the string scale 
$1/\sqrt{\alpha^\prime}\,$  string theory can be approximated by supergravity,
where dilatons and gravitons play an important role\cite{Po}. In particular we will
be interested in dilatons which are the string duals to scalar glueballs.
We consider  the dilaton to be in the $s-wave$ state, so we will not take
into account variations with respect to S$^5$ coordinates. 
The free dilaton field in AdS$_5$ with size $z_{max}\,$
can be cast into the form \cite{BB1}
\begin{eqnarray}
\label{QF}
\Phi(z,\vec x,t) &=& \sum_{p=1}^\infty \,
\int { d^3 k \over (2\pi)^{3}}\,
{z^{2} \,J_2 (u_p z ) \over z_{max}\,\, w_p(\vec k ) 
\,J_{3} (u_p z_{max} ) }\nonumber\\
&\times& \lbrace { {\bf a}_p(\vec k )\ }
 e^{-iw_p(\vec k ) t +i\vec k \cdot \vec x}\,
\,+\,\,h.c.\rbrace\,,
\end{eqnarray}

\noindent with $0\,\le\,z\,\le z_{max}\,$
and $w_p(\vec k ) \,=\,\sqrt{ u_p^2\,+\,{\vec k}^2}\,$ , $h.c.\,$ 
means hermitean conjugate and  
$u_p$ are defined by 
\begin{equation}
\label{up}
u_p z_{max}\,=\, \chi_{_{2\,,\,p}}
\end{equation}

\noindent such that the Bessel function satisfies $ J_2 (\chi_{_{2\,,\,p}} )=0$. 

The operators ${\bf a}_p\, ,\;{\bf a}^{\dagger}_p \,$ satisfy the commutation relations
\begin{equation}
\label{canonical1}
\Big[ {\bf a}_p(\vec k )\,,\,{\bf a}^\dagger_{p^\prime}({\vec k}^\prime  )
\Big]\,=\, 2\, (2\pi)^{3} w_p(\vec k )   
\delta_{p\,  p^\prime}\,\delta^{3} (\vec k -
{\vec k}^\prime )\,.
\end{equation}

On the boundary ($ z = 0)$ of the AdS slice we consider massive composite operators 
$\Theta_i(\vec x,t)\,$ representing glueballs with masses $\mu_i$. 
The algebra of the corresponding creation-annihilation operators can be written as
\begin{equation}
\label{canonical2}
\Big[ {\bf b}_i( \vec K )\,,\,{\bf b}_i^\dagger ({ \vec K }^\prime  )
\Big]\,=\, 2 (2\pi)^{3} \, w_i( \vec K ) \,\delta^3 ( \vec K -
{ \vec K}^\prime ) \,,
\end{equation}

\noindent where $ w_i(\vec K ) = \sqrt{ {\vec K}^2 + \mu^2_i}$.

In the previous work\cite{BB3} we considered a single glueball operator 
$\Theta(\vec x,t)$.
We have seen, following the general ideas of \cite{BB2},
 that the discretization introduced by considering a slice of AdS
makes it possible to establish a one to one mapping between bulk $(\vec k , u_p)$ 
and boundary ($\vec K$) momenta.
We assume a trivial mapping between the angular parts of $\vec k $ and
$\vec K  $. 
Then in order to find a relation between  $k = \vert \vec k\vert\, ,\, u_p$ 
and $K = \vert \vec K \vert$ we introduce 
a sequence of energy scales ${\cal E}_j$.
Defining $K_j$ to be a momentum in the interval ${\cal E}_{j-1} \le K \le {\cal E}_j\,$
we map the momentum space operator  ${\tilde \Theta}(K_j)\,$ 
in a one to one relation with the dilaton operator of momentum $k,u_p$:
$$ {\tilde \Theta}(K_j) \,\leftrightarrow {\tilde\Phi} (k,u_p) \,.$$
This way all the dilaton states are mapped into a single field on the boundary.
Note that each interval 
${\cal E}_{j-1} \le K \le {\cal E}_j\,$ is mapped into the entire range of 
$k$ corresponding to a fixed $u_p$. For each positive integer $j$ 
we choose a different $p$.

As we are looking at physical processes in the boundary theory 
which take place in a given energy range
we can take  ${\cal E}_1\,$  large enough  so that the first 
energy interval $0 \le K \le {\cal E}_1 \equiv {\cal E}\,$ contains all the relevant 
physics. Then only one interval for $K $ is necessary.
In this case the above mapping reduces to 
$$ {\tilde \Theta}(K_1) \,\leftrightarrow {\tilde \Phi} (k,u_p) 
\,\,\,\,\,(p \,\,{\rm fixed)}.$$

\noindent For simplicity we took $p=1\,$ in the previous work.

Here we want to describe physical processes involving a set of glueball operators
$\Theta_i(\vec x,t)\,\,\,(i=1,2, ...\,)\,$ using the same kind of mapping. 
If we again introduce momentum operators 
${\tilde \Theta}_i(K_j)\,$ with momentum $K_j$ in the interval 
${\cal E}_{j-1} \le K \le {\cal E}_j\,$ they would  not be mapped 
in a one to one relation with bulk operators ${\tilde \Phi} (k,u_p)$ 
unless $j$ is limited, since  $ i$ and $p$ are unlimited. 
Then such a mapping is possible if we introduce a restriction on the index $j$. 
The simplest choice is to take just one value for $j$.
This is obtained taking ${\cal E}_1\equiv {\cal E} $ 
large enough, which means that now $j = 1\,$. 
This recovers the previous solution and in this case 
the one to one mapping reads
$$ {\tilde \Theta}_i( K ) \,\leftrightarrow {\tilde \Phi} (k,u_i) ,$$

\noindent where we have dropped the index of $K_1$ since it is the only relevant
boundary momentum.  

This mapping can be written explicitly in terms of bulk and boundary 
creation-annihilation  operators. We will impose the same relation
proposed in \cite{BB3} 

\begin{eqnarray}
\label{ab}
k\,{\bf a}_i ( \vec  k ) 
&=& K \,{\bf b}_i ( \vec K  ) \nonumber\\
k\,{\bf a}^\dagger_i ( \vec k ) 
&=& K\,{\bf b}_i^\dagger ( \vec K  )\,.
\end{eqnarray}

\noindent For a general relation between bulk and boundary creation-annihilation
operators in AdS$_{n+1}$ see \cite{BB2}.

Requiring  that equations (\ref{ab}) preserve the canonical commutation relations 
(\ref{canonical1},\ref{canonical2}) one finds that the moduli of the momenta 
are related, for each bulk and boundary states by
\begin{equation}
\label{completo}
k = {u_i \over 2} \,\Big[ \,{ {\cal E}  +\sqrt{{\cal E}^2 + \mu_i^2 } 
			\over  K + \sqrt{K^2 + \mu_i^2}}
- { K + \sqrt{K^2 + \mu_i^2}\over {\cal E}  +\sqrt{{\cal E}^2 + \mu_i^2 }}\,\Big]\,,
\end{equation}

\noindent where $ 0 \le K \le {\cal E}\,$.
Note that this mapping contains as a particular case the previous one\cite{BB3}
which can be reobtained here if we consider the masses $\mu_i$ to be completely 
degenerated.

Now we associate the size $z_{max}$ of the AdS space with the mass of the lightest
glueball which we choose to be $\mu_1$  
\begin{equation}
z_{max} \,=\, {\chi_{_{2\,,\,1}}\over \mu_1}\,,
\end{equation}
 
\noindent so that from equation (\ref{up}) we have 
\begin{equation}
u_i \,=\,{\chi_{2\,,\,i}\over \chi_{2\,,\,1}}\,\, \mu_1 .
\end{equation} 

An approximate expression for the mapping (\ref{completo}) can be obtained 
choosing appropriate energy scales.
We take ${\cal E}$ to be the string scale $ 1/ \sqrt{\alpha^\prime}$ assuming  that 
$K \ll {\cal E}$. 
On the other side we restrict 
the momenta $K$ associated with glueballs to be much larger than their masses $\mu_i$.
Then we have 
\begin{equation}
\label{region}
\mu_i \ll K \ll {1\over \sqrt{\alpha^\prime}}.
\end{equation}

\noindent In this regime the mapping (\ref{completo}) reduces to 
\begin{equation}
\label{Kk}
k \,\,\approx\,{ u_i \,  \over 2 \,\sqrt{\alpha^\prime} K}\,. 
\end{equation}

\noindent This approximate mapping gives a high energy scaling similar to QCD\cite{BB3}.
Using the conditions (\ref{region}) together with the above mapping 
we see that the bulk momenta satisfy
\begin{equation}
u_i \ll k \ll \,\Big( {u_i \over \mu_i} \Big)\,{1\over \sqrt{\alpha^\prime}} .
\end{equation}

Note that the supergravity approximation holds for $ k \ll 1/ \sqrt{\alpha^\prime}\,$.
So in order to keep this approximation valid for all glueball operators $\Theta_i$
the factor $u_i / \mu_i $ should be nearly constant. We then impose that
 
\begin{equation}
{u_i \over \mu_i }\,=\,constant\,\,.
\end{equation}

So the glueball masses are related to the zeros of the Bessel
functions by
\begin{equation}
\label{QCD4}
{ \mu_i\over \mu_1 }\,=\,{\chi_{2\,,\,i}\over \chi_{2\,,\,1}}\,\,.
\end{equation}

Using the values of these zeros one finds the ratio of the glueball masses
for the state $0^{++}$ and its excitations. We are using the conventional notation
for these states with spin zero and positive parity and charge conjugation.
In order to compare our results 
from bulk/boundary holographic mapping with 
those coming from lattice we adopt the mass of the first state as an input.
Our results are in good agreement with lattice\cite{MASSL,MASSL2} and AdS-Schwarzschild 
black hole supergravity  calculations as seen  in table I.   
It is interesting to mention that an approach to
estimate glueball masses in Yang Mills$^\ast$ from  a deformed AdS space was discussed
very recently in \cite{ACEP}.

We can generalize the above results to  AdS$_{n+1}$. 
In this case massless  bulk fields are expanded in terms of the Bessel 
functions $J_{n/2} $ and the mass ratios for the $n$ dimensional "glueballs"  
are given in terms of their zeros. 
In particular for AdS$_4$, where one expects to recover results from QCD$_3$,
we find 
\begin{equation}
\label{QCD3}
{ \mu_i\over \mu_1 }\,=\,{\chi_{3/2\,,\,i}\over \chi_{3/2\,,\,1}}\,\,.
\end{equation}
\noindent Using this relation we obtain the ratio of masses presented in table
II together with lattice and $AdS$-Schwarzschild black hole supergravity calculations.
The agreement here is also good.

\begin{widetext}

\begin{table}[htbp]
\centering
\begin{tabular}{l|ccc}
QCD$_4$ state & lattice, $N=3$ &
Supergravity & Bulk/boundary \\
 \hline
 $0^{++}$ & $1.61 \pm 0.15$   & 1.61 {\rm (input)} & 1.61 {\rm (input)} \\
 $0^{++*}$ &  2.8   & 2.38 & 2.64 \\
 $0^{++**}$ &   - & 3.11 & 3.64 \\
 $0^{++***}$ &  -  & 3.82 & 4.64\\
 $0^{++****}$ &  -  & 4.52 & 5.63\\
 $0^{++*****}$ &  -  & 5.21 & 6.62\\
\end{tabular}
\parbox{6in}{\caption{ Masses of the first few $0^{++}$ glueballs for QCD$_4$ 
with $SU(N)$ and $N=3$, in GeV, from lattice\cite{MASSL,MASSL2}, from AdS-Schwarzschild 
black hole supergravity\cite{MASSG} and our results from bulk/boundary
holographic mapping, eq. (\ref{QCD4}).}}
\end{table}

\begin{table}[htbp]
\centering
\begin{tabular}{l|cccc}
QCD$_3$ state & lattice, $N=3$ & lattice, $N\rightarrow \infty$ &
 Supergravity& Bulk/boundary \\
 \hline
 $0^{++}$ & $4.329 \pm 0.041$ & $4.065 \pm 0.055$ & 4.07 ({\rm input})
& 4.07 ({\rm input}) \\
 $0^{++*}$ & $6.52 \pm 0.09$ & $6.18 \pm 0.13$ & 7.02 & 7.00\\
 $0^{++**}$ & $8.23 \pm 0.17$ & $7.99 \pm 0.22$ & 9.92 & 9.88 \\
 $0^{++***}$ &  - & - & 12.80 & 12.74 \\
 $0^{++****}$ &  - & - & 15.67 & 15.60\\
 $0^{++*****}$ & -  & - & 18.54 & 18.45\\
\end{tabular}
\parbox{6in}{\caption{$0^{++}$ glueball masses for QCD$_3$ with $SU(N)\,$
from lattice\cite{MASSL,MASSL2}(in units of string tension) , from AdS-Schwarzschild 
black hole supergravity \cite{MASSG} and our results from bulk/boundary
holographic mapping, eq. (\ref{QCD3}).  }}
\end{table}

\end{widetext}   

It is interesting to see if the AdS slice considered here can be 
related to the AdS-Schwarzschild black hole metric proposed by Witten\cite{Wi2}.
Witten's proposal for the case of QCD$_3$ corresponds to the ten dimensional 
metric\cite{MASSG}

\begin{eqnarray}
ds^2 &=& R^2 \Big( \rho^2 - {b^4\over \rho^2} \Big)^{-1} d\rho^2
+  R^2 \Big( \rho^2 - {b^4\over \rho^2} \Big) d\tau^2 \nonumber\\
&+& R^2 \rho^2 
(d\vec x)^2\,
+ \,R^2 d\Omega_5^2 \,\,,
 \end{eqnarray}

\noindent where $\rho \ge b\,$, $\,R^2 = l_s^2 \sqrt{4\pi g_s N} $ and $b $ is inversely 
proportional to the compactification radius of $S_1$ where the $\tau$ variable
is defined. 

If we qualitatively neglect the $\tau$  contribution to the 
metric in the limit of very little compactification radius
and then take the limit $\rho \gg b$ this metric is approximated by

\begin{equation}
ds^2 = { R^2 \over \rho^2}  d\rho^2
+  R^2  \rho^2 d\tau^2 
+ R^2 \rho^2 
(d\vec x)^2\,
+ \,R^2 d\Omega_5^2 \,\,.
 \end{equation}

\noindent That is an AdS$_4\times S^5\,$ space that takes a form similar
to eq. (\ref{metric}) if we
change the axial coordinate to $z\,=\,1/\rho $. 
In Witten's framework one must impose regularity conditions
at $\rho = b$ because of the presence of the horizon at this position.
In our approximation in order to retain this physical
condition we impose boundary conditions there 
and associate it to the cut of our slice ($b = 1/z_{max}$).
This AdS$_4$ slice is the one used to estimate the glueball mass ratios 
related to the three dimensional gauge theory (\ref{QCD3}). 
So we can think of our AdS$_4$  slice as a naive approximation 
to Witten's proposal.

An analogous situation could also be considered for the 
Witten's proposal to QCD$_4$. In that case the situation is more involved
because of the form of the metric coming from the compactification of 
AdS$_7\times S^4$. 

In conclusion we have seen that the bulk/boundary holographic mapping which reproduces
the high energy scaling of QCD like theories can also be applied to estimate 
glueball mass ratios. 
We hope that this mapping can be used  to describe other particle states
that may be related to some properties of QCD.

It is important to remark that one can obtain a similar result for the 
ratio of the glueball masses considering other mappings between bulk and boundary 
creation-annihilation operators instead of eq. (\ref{ab}).
For example one could take ${\bf a}_i \,= \,{\bf b}_i\,$. This would contain the solution 
$\,k = K$  implying that the masses of the glueballs are identically equal 
to the values of axial bulk momenta $u_i$. However such a trivial mapping does not 
seem to reproduce the high energy QCD scaling.

Let us mention that we used a solution for the dilaton field corresponding to
Dirichlet  boundary conditions at $z = 0 $ and $ z= z_{max}$.
This allows the existence of Bessel functions but not the divergent Neumann solutions. 
Other boundary conditions can also be considered in the same context.  
 
We have also obtained these mass ratios for scalar glueballs starting with
the same AdS slice as discussed here without using the holographic mapping 
of ref \cite{BB3} but assuming the stronger condition of relating directly the dilaton
modes with the glueball masses\cite{BB4}. The consistency between these results 
seems to indicate that the holographic mapping found before may indeed be valid
within the approximations and the energy region considered. 

  What is surprising in this bulk/boundary holographic mapping is that it seems 
to describe features of both high and low energy regimes of the boundary theory
since it gives information about the high energy scaling and mass spectrum. 

\noindent {\bf Acknowledgments.} We would like to thank J. R. T. Mello Neto 
for important discussions. The authors are partially supported by CNPq, FINEP, 
CAPES and FAPERJ 
- Brazilian research agencies.



\end{document}